\documentclass[aps,twocolumn,pra,tightenlines,floatfix,showpacs,superscriptaddress]{revtex4-1}
\usepackage{amssymb}
\usepackage{amsmath}
\usepackage{mathrsfs}
\usepackage{graphicx}
\usepackage{array}

\newcommand{\bs}[1]{\boldsymbol{#1}}

\newcommand{\bckt}[1]{\left[ #1 \right]}
\newcommand{\abs}[1]{\left| #1 \right|}

\newcommand{\upa}[0]{\uparrow}
\newcommand{\dna}[0]{\downarrow}

\newcommand{\mrm}[1]{{\mathrm{#1}}}

\newcommand{\meq}[1]{\begin{equation} #1 \end{equation}}
\newcommand{\mea}[1]{\begin{align} #1 \end{align}}

\begin{document}

\title{Pseudogap Effects of Fermi Gases in the Presence of
A Strong Effective Magnetic Field}

\author{Peter Scherpelz}
\author{Dan Wulin}
\author{K. Levin}
\affiliation{James Franck Institute and Department of Physics,
University of Chicago, Chicago, Illinois 60637, USA}
\author{A. K. Rajagopal}
\affiliation{Inspire Institute Inc., Alexandria, Virginia 22303, USA}
\affiliation{Harish-Chandra Research Institute, Chhatnag Road, Jhunsi,
Allahabad, 211019, India}

\date{\today}

\begin{abstract}
We address the important question of how to characterize the normal state
of fermionic superfluids
under the influence of a strong effective magnetic field, implemented through
rapid rotation or novel artificial field techniques. We consider the
effects of crossing from BCS to BEC and the role of non-condensed pairs, or
pseudogap effects.
Using
a simple extension of
Gor'kov theory we demonstrate how these pairs organize above the
transition $T_c$
into precursors of a vortex configuration, which  
are associated with distortions
of the ideal Abrikosov lattice. This non-uniform normal state appears to enable
``Bose condensation" in a field which is otherwise problematic due to
the effective one-dimensionality of Landau level dispersion.
\end{abstract}

\maketitle

The role of non-condensed pairs in fermionic and bosonic superfluids
under the influence of a strong effective magnetic field has not been
clearly addressed in the literature. 
One could imagine that in anticipation of the vortex 
configuration
of the superfluid phase, these non-condensed pairs exhibit some
degree of inhomogeneity. 
The goal of the present paper is to investigate this normal phase.
Our study should lead to a reconsideration of previous work on
rotating cold Fermi gases
\cite{Horotation,*Radzihovskyrotation,*Cooperrotation}
where the upper critical rotation frequency was computed under the assumption
that non-condensed pairs were not present, even in the BEC regime.
It may also be relevant to ``normal state vortex" models \cite{Ong2,anderson_2007}
which are argued to be important in high temperature superconductivity.  
Furthermore, we note the community excitement over
proposals
to establish strong effective magnetic fields through ``artificial'' means.
These are implemented through
either asymmetric tunneling in a lattice \cite{kolovsky_2011}
or geometric gauge potentials \cite{dalibard_2011,cooper_2011}.
This makes our work on the interplay of rapid rotation and superfluidity
particularly topical.

Finally, we believe our work bears on 
a  puzzling aspect of superconductivity in high magnetic fields, even within the
BCS framework. As has been emphasized in the literature, 
the degeneracy of Landau levels implies that the fluctuations around the
BCS phase are 
effectively one-dimensional \cite{lee_1972}, representing
the free propagation of particles along
the field direction; this 
is well known \cite{schafroth_1955,*ullah_1991} 
to be problematic for stable
superconductivity.

In this paper we suggest that the
origin of this latter difficulty is due to the fact
that in BCS theory, Cooper pairs only exist in a condensed state.
Here we 
show that by introducing non-condensed pairs associated with
a finite pairing gap 
(or ``pseudogap") at the onset of
condensation, a strict ``dimensional reduction''\cite{lee_1972} is no longer
present; three-dimensional behavior leading to stable condensation
at $T_c$ can then occur.
Our approach is to be contrasted with previous attempts
\cite{alexandrov_1987,*alexandrov_1993}
to address the
early concerns raised by Schafroth \cite{schafroth_1955} in Bose gases.
Here we emphasize 
the contrast with
BCS theory, where the only pairs under consideration are in the
condensate.

One goal
of this paper is to stimulate experimental searches for
the predicted precursors of the vortex configurations below $T_c$.
Here we characterize the degrees of freedom associated with
non-condensed pairs and demonstrate that
they create pair density inhomogeneities.
Although these
excited pairs become progressively more important as the superconductor
crosses from BCS to BEC (and pairing becomes more stable), they may well
be
necessary even in the BCS limit for stable condensation.  These ideas
relate to earlier work on pair density waves 
\cite{tesanovic_1994}, but are in contrast
to previous studies on the interplay of a pseudogap and magnetic field
\cite{kao_2001,*pieri_2002}
where density inhomogeneities were not contemplated.

To understand these non-condensed pairs we are guided by
Landau-Ginzburg (LG) theory where in zero magnetic field
non-condensed pairs are associated with finite center of
mass momentum $\bs q$; these
represent gapless excitations as $\bs q \rightarrow 0$.
In non-zero field, 
the natural counterpart should be associated with
slightly distorted configurations of the 
vortices.
These pair excitations become gapless as they approach the
lowest energy superconducting vortex
configuration. However, we also emphasize that these non-condensed pairs are
distinct from previously considered Landau-Ginzburg vortex lattice fluctuations 
which only address the
condensate \cite{maniv_2001,rosenstein_2010}.
We now proceed to describe how we incorporate these non-condensed
pairs, and their effects at and above the superfluid transition.

\textbf{Rewriting the Gor'kov Equations}
We begin with a BCS-type theory, 
and proceed to separate out the effects of pairing
and condensation by introducing non-condensed pairs appropriate for high
effective magnetic fields. 
First, we introduce a Landau level
representation for the Gor'kov equations.
The
real-space Gor'kov coupled equations for the 
gap $\Delta(\bs r)$ and the
fermionic Green's function
$G(\bs r,\bs r'; i\omega)$ are:
\mea{
G(&\bs r,\bs r'; i\omega)
= G^0(\bs r,\bs r'; i\omega) - \int d\bs r'' d \bs r''' G^0(\bs r,\bs
r'';i\omega) \nonumber \\
&{} \times \Delta(\bs r'')G^0(\bs r''',\bs r''; -i\omega) \Delta^\dag(\bs
r''')G(\bs r''',\bs r'; i\omega) \label{GorkovGR}}
\meq{\Delta^\dag(\bs r) = \frac{g}{\beta}\sum_{i\omega}\int d\bs r'G(\bs r',\bs
r;i\omega)G^0(\bs r',\bs r; -i\omega)\Delta^\dag(\bs r')\label{GorkovGapR}}
Throughout this paper $i\omega$
($i\Omega$) will be used to denote discrete fermionic (bosonic) Matsubara
frequencies, with the traditional subscripts omitted for clarity.
Introducing a Landau level basis for the fermions indexed by $m =
(N,p,k_z)$ where $N$ is the Landau level, $p$ the degenerate Landau level index,
and $k_z$ the momentum parallel to the magnetic field, we write the bare
Green's function 
$G^0(\bs r,\bs r';i\omega) = \sum_n \psi_n(\bs
r)\psi_n^\dag(\bs r')/(i\omega-\xi_n)$
where $\xi_n$ is the single-particle
energy; the dressed Green's function is
$G(\bs r,\bs r';i\omega) =
\sum_{mm'}G_{mm'}(i\omega)\psi_m(\bs r)\psi^\dag_{m'}(\bs
r')$ \cite{vavilov_1997}.
The self
energy,
given by
$\Sigma(\bs r,\bs r';i\omega) = -\Delta(\bs r)\Delta^\dag(\bs r ')G_0(\bs r',\bs
r;-i\omega),$
is rewritten as
$\Sigma_{mm'}(i\omega) =
-\sum_nG^0_n(-i\omega)\Delta_{mn}\Delta^\dag_{m'n},$ and the number equation
necessary for a self-consistent solution is $N = \frac{2}{\beta}
\sum_{m,i\omega} G_{mm}(i\omega)$.
Defining
$\Delta_{mn} \equiv \int d\bs r \Delta(\bs r)\psi_m^\dag(\bs
r)\psi_n^\dag(\bs r)$, and integrating over position variables yields
\mea{G_{mm'}(i\omega) &= G^0_m(i\omega)\delta_{m m'} \nonumber \\
&{} - \sum_{l n }
G^0_{m}(i\omega)\Delta_{ml}G^0_l(-i\omega)\Delta^\dag_{l n}
G_{nm'}(i\omega) .\label{eq:SSGF}}
\meq{1 = \frac{g}{\beta}
\sum_{i\omega}\sum_{m m' n}\frac{\Delta_{m' n}\Delta^\dag_{mn}}{\int d\bs r
\abs{\Delta(\bs r)}^2}G_{mm'}(i\omega)G^0_n(-i\omega).\label{eq:SSGap}} These
correspond to 
Eqs.~(\ref{GorkovGR})-(\ref{GorkovGapR}) respectively in the Landau level 
basis. 
Eq.~\eqref{eq:SSGap} is equivalent to that found elsewhere in
Refs.~\cite{ryan_1993,dukan_1994}. As in these references
we restrict our consideration to intra-Landau level pairing. 
This assumption is important for arriving at a
tractable scheme and
a good approximation in the high-field regime \cite{tesanovic_1998}. 

It is useful to recognize that
the nonlinear gap equation, Eq.~\eqref{eq:SSGap},
applies to all $ T \leq T_c$. 
This nonlinearity is reflected in the presence of one dressed and one bare
Green's function, as opposed to the two bare Green's functions associated with
the instability onset in strict BCS theory.
This gap equation can be thought of as a Bose-Einstein
condensation condition \cite{chen_2005}
which reflects the vanishing chemical potential of the pairs
below $T_c$, and which becomes finite above.

\textbf{Characterizing Non-condensed Pairs} 
Outside of the weak-coupling limit, pairs may form in kinetically excited states
\cite{chen_2005}.  These pairs appear above the superfluid transition
temperature $T_c$, corresponding to a pseudogap phase, and should also persist
below the transition. 
To include these non-condensed pairs, we must
therefore introduce variables corresponding to their excitation parameters.

In the $z$-direction parallel to the field, 
we introduce a total momentum $q_z = k_{z_1} + k_{z_2}$ of
the pair, in analogy with the zero-field case \cite{chen_2005}.  Perpendicular
to the effective magnetic field, however, both condensed and non-condensed pairs
may lie in the same Landau level.  We posit, therefore, that non-condensed pairs
correspond to different real-space gap configurations $\Delta(\bs r)$, which we
now parametrize by $\zeta$.  We single out
$\zeta_0$ which denotes the condensate gap configuration. In
general non-condensed pairs can occupy other functional forms of
$\Delta(\bs r)$, for which the parameter $\zeta$ will generally be 
close to $\zeta_0$, associated with low-energy excitations as described below.

\begin{figure}
\begin{tabular}{m{0.84in}m{4in}}
\includegraphics[clip = true, scale = 0.27]{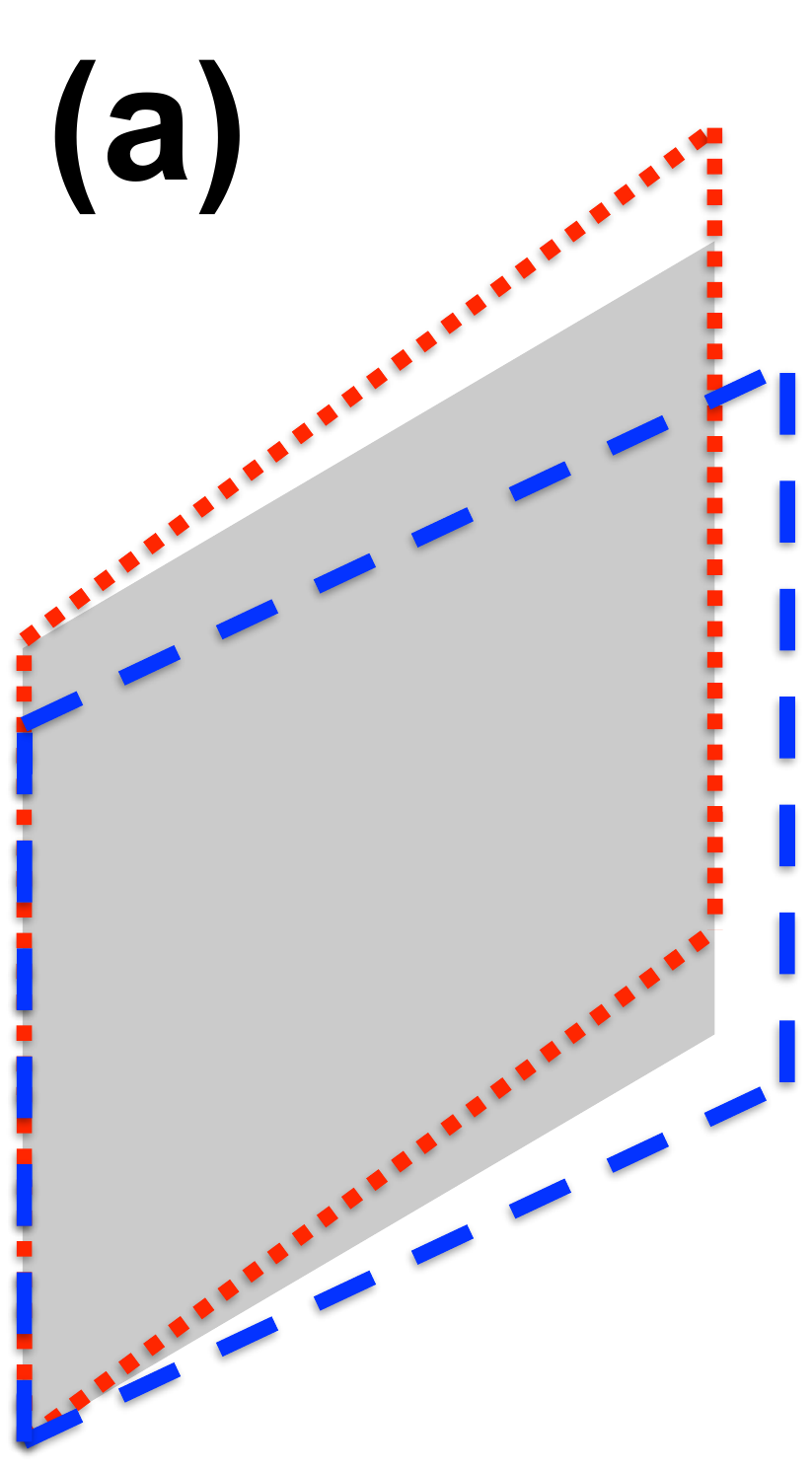} &
\includegraphics[clip = true, scale = 0.35]{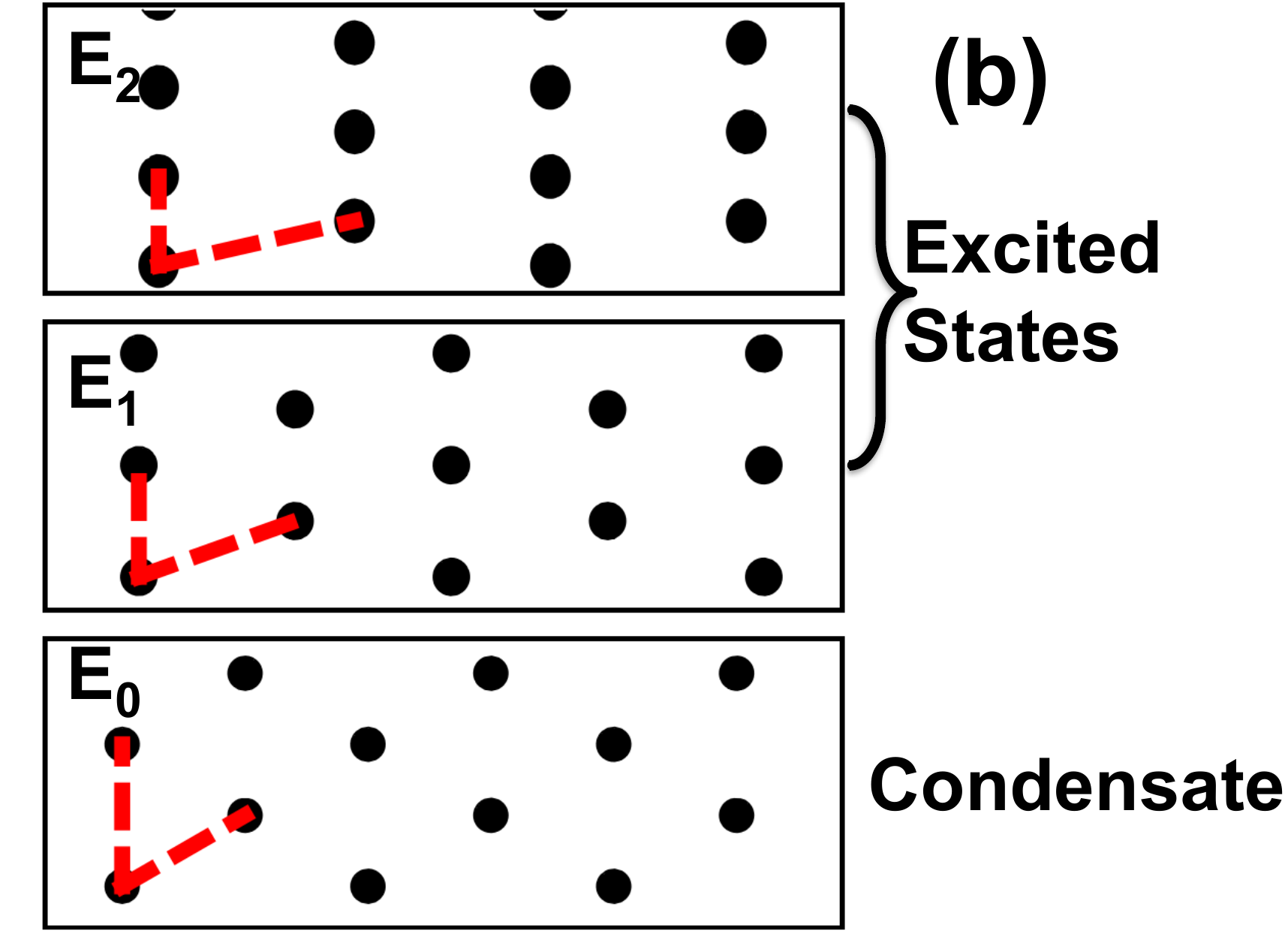}
\end{tabular}
\includegraphics[scale = 0.45]{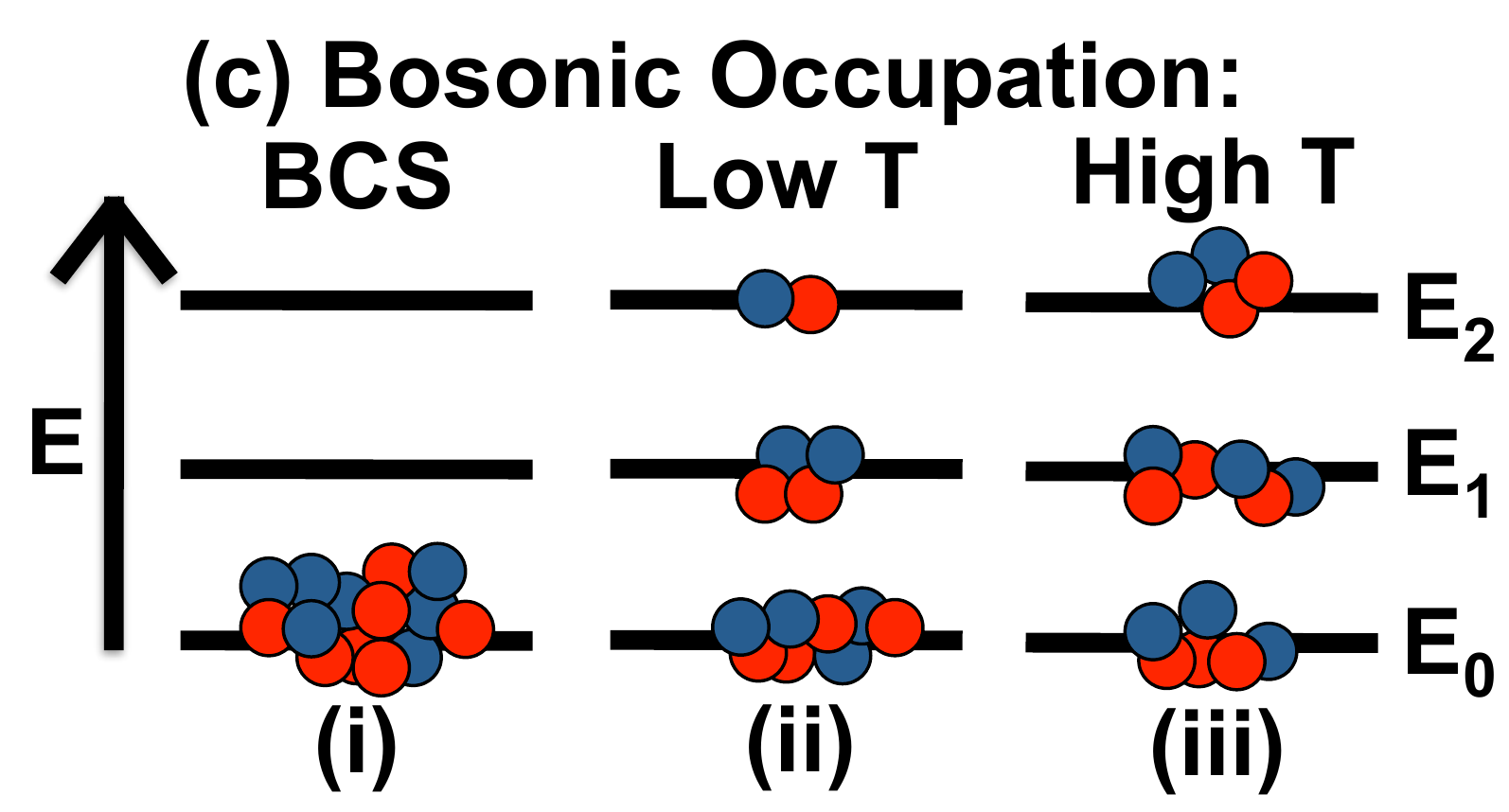}
\caption{\label{fig:Distortions} (Color online) A diagram of the different
distortions of the Abrikosov lattice, the resulting splitting in energy
levels, and the different occupation statistics that result.  
\textbf{(a)}: The shaded gray unit cell is the optimal
lattice configuration, whereas the red dotted unit cell corresponds to an
excitation of $b_y/a$ while the blue dashed unit
cell corresponds to an excitation of $b_x/a$.  \textbf{(b)}: Real-space diagrams
of three different values for $b_x/a$, 
showing zeroes of $\Delta^0(\bs r)$ for each
configuration (black circles) and lattice vectors (dashed red lines).  $E_0$ is
the optimal configuration, while $E_1$ and $E_2$ 
are progressively higher in energy.  \textbf{(c)}: The pairs (here pairs of
blue (spin up) and red (spin down) fermions) are now able to
occupy a continuum of energy levels corresponding to different lattice
configurations.  For the three displayed configurations,
\textbf{(i)}: The BCS approach results in only the optimal
configuration being occupied (also the case in this system at $T = 0$).
In contrast, in this system for $T \neq 0$, higher energy
levels can also be occupied by pairs. \textbf{(ii)}: An example of
occupation statistics at a low temperature -- most pairs are in $E_0$.
\textbf{(iii)}: At a higher temperature, more pairs are in excited states.
}
\end{figure}

In our mean-field approach, we consider the condensate function
$\Delta(\bs r,\zeta_0)$ to be that of the optimal triangular lattice.
Non-condensed pairs can then occupy a two-dimensional continuum of other 
Abrikosov lattice configurations.  Mathematically, 
we use the Landau gauge $\bs A = (0,H\hat{\bs x},0)$ and an Abrikosov
lattice with unit vectors $\bs a = (0,a,0)$ and $\bs b = (b_x,b_y,0)$, where
$ab_x = \pi l_H^2$ with $l_H = \sqrt{\hbar c/eH}$ the magnetic Hall length.
The two clear mean-field distortions available to excited pairs are 
associated with
changing
$b_x/a$, and changing $b_y/a$ (see
Fig.~\ref{fig:Distortions}) \cite{saint-james_1969}.
Both of these are higher in energy relative to 
the optimal Abrikosov lattice. We then follow Ref.~\cite{saint-james_1969}
in setting $\zeta = b_y/a + ib_x/a$, for which the optimal triangular lattice 
condensate 
configuration is $\zeta_0 = 1/2+i\sqrt{3}/2$. 
To transform to state-space, we can associate with 
each distortion a normalized real-space gap 
configuration $\Delta^0(\bs r,\zeta)$, where $\int d\bs r
\abs{\Delta^0(\bs r,\zeta)}^2 = 1$, from which $\Delta^0_{mn}(\zeta) \equiv
 \int d\bs
r \Delta^0(\bs r,\zeta) \psi_m^\dag(\bs r)\psi_n^\dag(\bs r)$ can also be
calculated.  

In summary, 
these non-condensed pairs are now specified by three degrees of freedom,
thereby compensating the problematic dimensional
reduction and restoring the possibility of stable condensation. We can also use
these parameters to rewrite the gap equation, Eq.~\eqref{eq:SSGap}, 
by defining a pair susceptibility 
$\chi(\zeta,q_z;i\Omega) \equiv$
\meq{
\frac{1}{\beta}\sum_{i\omega,m,m'}
\phi_{mm'}^2(\zeta) G_{mm'}(i\omega) G^0_{N} (q_z-k_z;i\Omega-i\omega)
\label{eq:chi}}
so that Eq.~\eqref{eq:SSGap}
assumes a simple and suggestive form
\begin{equation}
1+g\chi(\zeta_0,0;0) = 0.
\label{eq:4b}
\end{equation} 
In
Eq.~\eqref{eq:chi}, 
$G^0$ is written in terms of the Landau level $N$ and 
$z$-momentum of $m =
(N,p,k_z)$ (with $N_m = N_{m'}$ and $k_{z_m} = k_{z_{m'}}$) and
$\phi^2_{mm'}(\zeta) = \sum_{n}\Delta^0_{mn}(\zeta)\Delta^{0\dag}_{nm'}(\zeta).$

\begin{figure*}[tbp]
\includegraphics[scale = 0.80, clip = true, trim = 0.15in 0.1in .7in .95in]{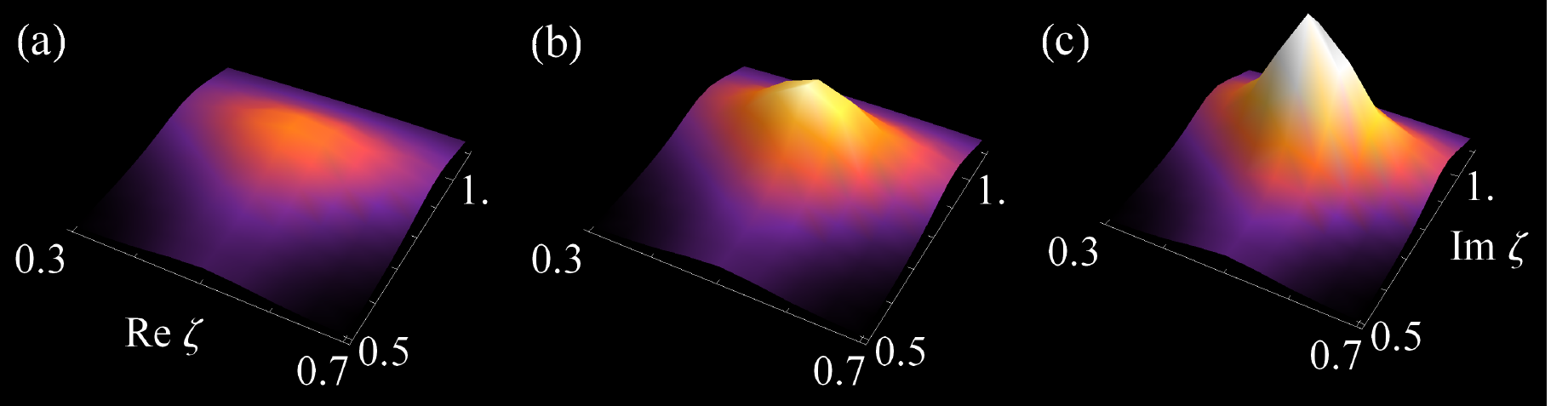}
\includegraphics[scale = 0.15]{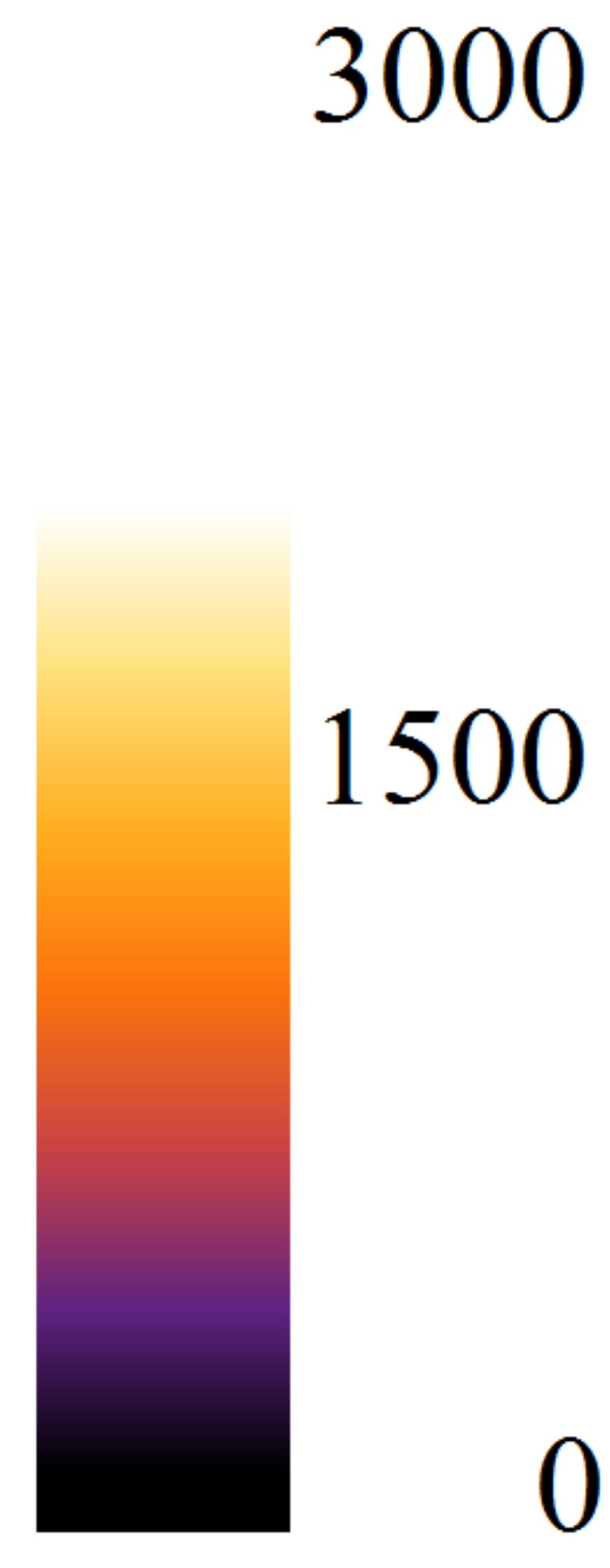}
\caption{\label{fig:tResults} (Color online)
A three-dimensional plot of
$\abs{t^\mrm{pg}(\zeta,0;0)}$ against lattice configurations $\zeta$. 
These plots are
slightly above $T_c$,
such that in the rightmost plot 
$\abs{g^{-1}} - \chi(0,0;0) =
-\mu_\mrm{pair}/Z_0$ where $\mu_\mrm{pair}/Z_0$ is chosen to be $-3\times10^{-4}$ (see
Ref.~\cite{chen_2005} for a description of these parameters), and
the other two plots share the same interaction $g$ and fermionic chemical
potential $\mu = 1.0$, but different $(\Delta^\mrm{pg},T)$: \textbf{(a)} 
$(5.75,1.1118);$ \textbf{(b)} $(5.90,1.0480);$ \textbf{(c)} $(6.00,1.00).$ The
fermions lie in the lowest Landau level for simplicity.
}
\end{figure*}

\textbf{Green's Function of Non-condensed Pairs}

Equation \eqref{eq:4b} suggests that there is a $t$-matrix 
(or a summation of particle-particle ladder
diagrams), which is related to the pair susceptibility 
Eq.~\eqref{eq:chi}
and which diverges at and below $T_c$.
More precisely, we have shown in
Ref.~\cite{scherpelz_2012b}
that, not surprisingly, this 
$t$-matrix is given by 
\mea{
t^\mrm{pg}(\zeta,q_z;i\Omega) &\equiv
\frac{g}{1+g\chi(\zeta,q_z;i\Omega)} \label{eq:21}.}
Moreover, this $t$-matrix will lead to a self energy 
contribution \cite{scherpelz_2012b}
in parallel with what is
found in Gor'kov theory via the condensate t-matrix,
$\Sigma^\mrm{pg}_{mm'}(i\omega) = $
\meq{\frac{1}{\beta}\sum_{\zeta,q_z,i\Omega}
\phi^2_{mm'}(\zeta)t^\mrm{pg}(\zeta,q_z;i\Omega)
G^0_{N}(q_z-k_z;i\Omega-i\omega)
\label{eq:scpgse}. }
This self energy is then fed back into the Green's function which
enters into the pair susceptibility, and into the self-consistently
determined $t$-matrix.

Because of  
the mixing
through the self-energy of an infinite number of real-space gap configurations,
the calculation of the pair susceptibility
$\chi$ is not analytically tractable.
To make progress without the distraction of heavy
numerics, we approximate the pair susceptibility
as
$\chi(\zeta,q_z;i\Omega) \approx$
\meq{ \frac{1}{\beta}\sum_{i\omega}\sum_{m,m'}
\phi_{mm'}^2(\zeta) G_{mm'}(\zeta;i\omega) 
 G^0_{N} (q_z-k_z;i\Omega-i\omega)
}
where we have, in effect, decomposed $G_{mm'}$ into separate
contributions each associated with a distinct lattice structure.
In this way Eq.~\eqref{eq:scpgse}, which becomes
\mea {\Sigma^\mrm{pg}_{mm'}(\zeta;i\omega) &\approx
\frac{1}{\beta}\phi^2_{mm'}(\zeta)\sum_{\zeta',q_z,i\Omega}
t^\mrm{pg}(\zeta',q_z;i\Omega) \notag \\
&{}\times G^0_{N}(q_z-k_z;i\Omega-i\omega),}
determines the Green's functions
$G_{mm'}(\zeta;i\omega)$. 

Importantly, we can further write that 
$\Sigma_{mm'}^\mrm{pg}(\zeta;i\omega) \approx -\phi^2_{mm'}(\zeta)
\abs{\Delta^\mrm{pg}}^2
G^0_N(-k_z;-i\omega),$
where 
the gap contribution from the
non-condensed pairs is given by
$\abs{\Delta^\mrm{pg}}^2 \equiv -\frac{1}{\beta}\sum_{\zeta,q_z,i\Omega}
t(\zeta,q_z;i\Omega).$
To derive this simplified (BCS-like) expression for the pseudogap
self energy in a strong ``magnetic field" 
we have used the fact that the small chemical potential of the pairs 
implies that $t^\mrm{pg}$
is strongly peaked around $(q_z,
i\Omega)=(0,0)$ near and below $T_c$.
When we consider a diagonal pairing scheme \cite{dukan_1994},
the Green's
function that results has a familiar BCS form
$G_{mm}(\zeta,i\omega) =
(i\omega + \xi)/\bckt{(i\omega)^2-\xi^2-
\abs{\Delta^\mrm{pg}}^2\phi_{mm}^2(\zeta)}$.

In effect,  the main approximation we have made is to consider 
each vortex configuration as an
``independent system,'' sharing self-consistently the same magnitude of the
energy gap $\abs{\Delta^\mrm{pg}}^2$,
but possessing a distinct
real-space gap parameter $\Delta^0(\bs r,\zeta)$ which in turn provides a unique
form factor $\phi^2_{mm'}(\zeta)$ in the self-energy.
 This is similar in approach to considering an LG energy
functional for the different forms of the gap parameter. In contrast
to LG theory,
here we explicitly incorporate
the fermionic nature of the pairing.

\textbf{Fermionic Constituents of the Pairs}
This discussion has been cast in terms of the basis
eigenstates
$\psi_m(\bs r)$
which we now need to specify.
Here we use the magnetic translation
group basis 
\cite{dukan_1991,dukan_1994,akera_1991,*norman_1995,*nicopoulos_1991}, 
although we have also explored an alternative orbit-center
based pairing basis \cite{ryan_1993,scherpelz_2012b}.
Magnetic translation group (MTG) pairing 
relates to the
Abrikosov lattice and is associated with a Bloch-like index $\bs k =
(k_x, k_y)$.
The unit cell for the MTG 
has unit vectors $2\bs a$ and $\bs b$, which
means in turn that the state basis is dependent on $\zeta$.  Pairing
for the $\zeta$ appropriate to the basis occurs 
with a single partner
between opposite $\bs k$ \cite{dukan_1994}, 
i.e. $\Psi^\mrm{pair}_{N,\bs k,q_z}(\bs r) = 
\psi^\mrm{fermion}_{N,\bs
k,k_z,\upa}(\bs r)\psi^\mrm{fermion}_{N,-\bs k,-k_z+q_z,\dna}(\bs r)$.
$\Delta^0_{mn}(\zeta)$ for these pairs is computed elsewhere
\cite{dukan_1994,scherpelz_2012b}.

\begin{figure}[tbp]
\includegraphics[clip = true, trim = 0.0in 0.0in 0.0in 0.0in, scale = .46]{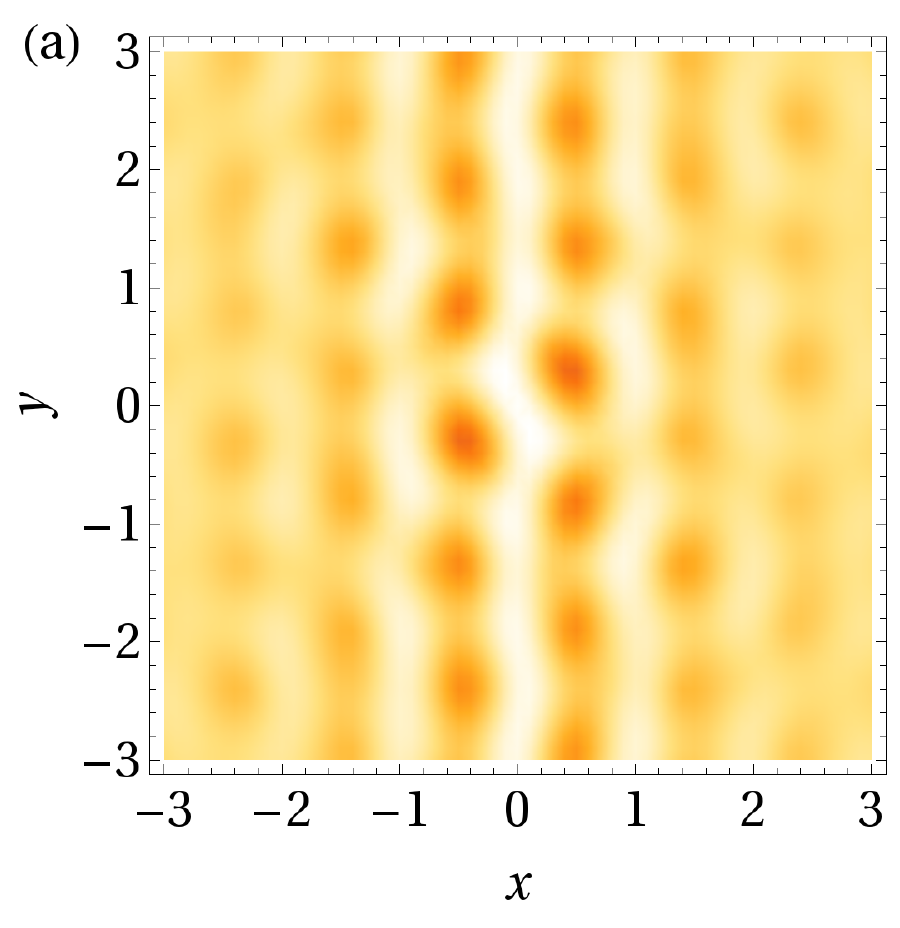}
\includegraphics[clip = true, trim = 0.0in 0.0in 0.0in 0.0in, scale = .45]{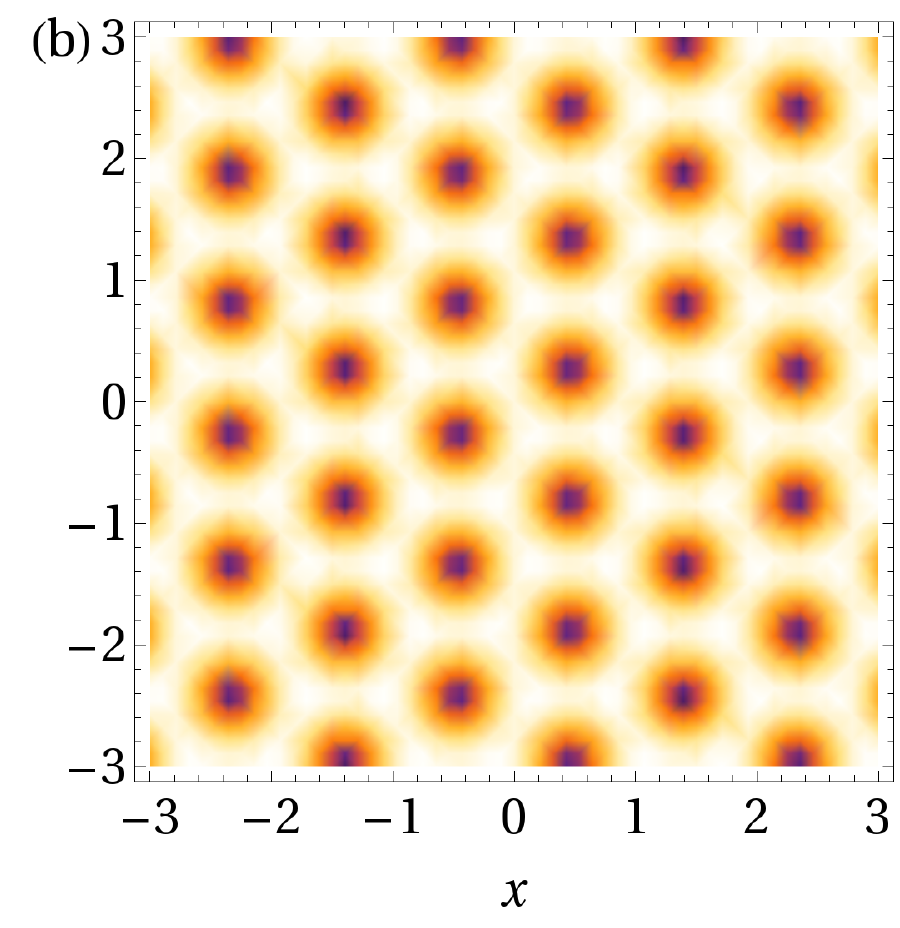}
\includegraphics[scale = 0.13, clip = true, trim = 0 .1in 0 1.5in]{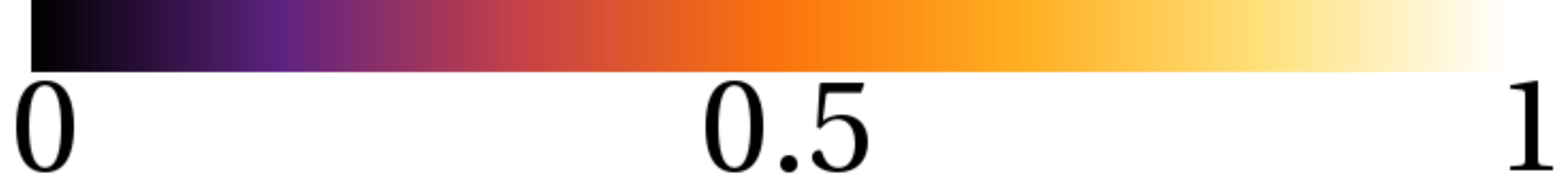}
\caption{\label{fig:cResults} (Color online)
\textbf{(a)}: A density plot of the total
squared energy gap $\abs{\Delta^\mrm{pg}}^2$ in real space, corresponding to
$\mu_\mrm{pair}/Z_0 = -3\times10^{-4},\ \mu = 1.0,\  
\Delta^\mrm{pg} = 6.0,$ and $T = 1.0$
(the same parameters as in Fig.~\ref{fig:tResults}c).
This calculation is done through a discrete sampling of 1,360 points in
$\zeta$-space, and normalized to the largest value of the gap.
\textbf{(b)}: A density plot of the
condensate energy gap $\abs{\Delta(\bs r,\zeta_0)}^2$ only, which corresponds
to the total energy gap at zero temperature, again normalized to its largest
value.
}
\end{figure}

\textbf{Results} With this formalism in place, we are in a position to examine the
underlying physics of
how superfluidity with pre-formed pairs takes place
in the presence of a high effective magnetic field. 
This
addresses 
the difficulty
that a stable transition seems to require some
inhomogeneity in the normal state
\cite{schafroth_1955,*ullah_1991}.
In
Fig.~\ref{fig:tResults} we
plot the $t$-matrix vs.~lattice configuration
$\zeta$ for three different sets of effective
temperatures,
demonstrating that as condensation is approached from higher temperatures,
the occupation of lattice states 
near 
the ideal triangular Abrikosov lattice 
($\zeta = \zeta_0$) begins to peak. 
Precisely at
$T = T_c$, a delta
function results at $\zeta = \zeta_0$. Nevertheless at the
transition 
there is still considerable weight associated with other lattice configurations
reflecting the fact that
the pseudogap
$\abs{\Delta^\mrm{pg}}^2$ remains finite.
The condensate contribution corresponds to
a perfect triangular lattice 
which necessarily has small weight near $T_c$.

Of particular interest is the 
real space reflection of
these distorted Abrikosov lattice 
contributions. To illustrate this precursor
vortex configuration we evaluate
\meq{\abs{\Delta^\mrm{pg}(\bs r)}^2 =\frac{1}{\beta}
\sum_{\zeta,\bs q_z,i\Omega}
t^\mrm{pg}(\zeta,q_z;i\Omega) \abs{\Delta^0(\bs r,\zeta)}^2,}
which is a weighted average of the gap (squared). This 
is compared with the counterpart for a fully
condensed system
in Fig.~\ref{fig:cResults}.
By addressing the square of the gap, we emphasize 
that there is no phase information in the normal state
pseudogap. It should be noted that the point
$\bs r = 0$ is chosen as a point of ``symmetry breaking'' or pinning center
which breaks the translational symmetry available in the selection of each
$\Delta^0(\bs r,\zeta)$.

\textbf{Conclusions} 
In this paper we have addressed
an important and in principle 
testable prediction in the cold Fermi gases: the
presence of precursor vortex configurations in the normal (pseudogap)
phase. 
Indeed, the role of non-condensed bosons in rapidly rotating condensates
has not been elucidated even for the atomic Bose gases. Here, too,
one might expect a precursor vortex 
configuration.
We have, moreover,
elucidated
the nature and role of excited pair states throughout the BCS-BEC crossover 
in high effective magnetic fields, showing that these
are associated with
distortions of the Abrikosov lattice.

The concept of
a ``normal state vortex" liquid
is also rather widely discussed in the context of high $T_c$
superconductors \cite{Ong2,anderson_2007}.
There are, of course, a host of controversial issues associated
with the cuprate pseudogap, but
Fig.~\ref{fig:cResults}
illustrates one scenario for
how one might think about this phenomenon in the
context of pre-formed pairs in a strong magnetic field.

The experimental 
implementation of this work in cold gases
will depend on reaching a regime in which the
Landau level spacing is much larger than the gap, 
although we expect the qualitative aspects regarding
the role of pair density inhomogeneities to apply to much lower fields. 
While rapid rotation may be able to reach
this regime, current proposals for artificial fields 
\cite{kolovsky_2011,dalibard_2011,cooper_2011} 
also show promise. 
In this paper, two spin states which interact via a Feshbach
resonance must both feel the same effective field, which may make geometric
gauge-based proposals more challenging. Regardless of the implementation, 
reaching this regime would facilitate a large body
of new research into BCS-BEC phenomena.

We are grateful to Jonathan Simon for helpful discussions. 
This work is supported by NSF-MRSEC Grant 0820054. P.S.~acknowledges 
support from the Hertz Foundation.

\bibliography{Review3,Review4}

\end{document}